\documentstyle[twoside,fleqn,espcrc2,epsf]{article}
\title{\bf Lattice Perturbation Theory}
\author{Colin~J.~Morningstar
\address{Department of Physics \& Astronomy, University of Edinburgh,
  Edinburgh EH9 3JZ, Scotland}}
\begin{document}

\begin{abstract}
Sources of uncertainties in perturbative calculations, tadpole
improvement and its role in lattice perturbation theory, and
six recent calculations are discussed.
\end{abstract}

\maketitle
\section{INTRODUCTION}
Perturbation theory (PT) appears in many important theoretical and
practical roles in lattice field theory.
Conceptual issues related to the continuum limit can be studied
in a perturbative framework, such as the renormalizability of lattice
gauge theories \cite{reisz}.  Perturbative calculations are used
in relating lattice quantities to those defined in continuum schemes
and in constructing Symanzik-improved lattice actions and operators.
A recent determination of the strong coupling constant \cite{alphas}
and a new construction of lattice chiral fermions \cite{chiral}
relied on perturbative expansions.  PT often helps us to better
understand our numerical computations and provides useful checks.

Lattice perturbation theory is a very broad topic.  Here, I
focus on weak-coupling PT in zero-temperature lattice QCD.
First, two important sources of uncertainties in lattice PT
calculations are briefly discussed. Next, tadpole improvement
and its role in lattice PT are described.  I then survey
six selected lattice PT papers or series of papers
from the recent past.  Developments concerning
renormalons are omitted, as these will be presented in the
talk by Sachrajda \cite{renormalon}.

\section{SOURCES OF UNCERTAINTIES}
In this section, I briefly discuss two important sources of
uncertainties in any calculation which
relies on lattice perturbation theory; namely, the very
applicability of PT and the choice of expansion
parameter in finite-order approximations.  Of course, these issues
must also be confronted in continuum perturbation theory.

\subsection{Applicability}
Because of asymptotic freedom, one generally expects that quantities
in QCD dominated by momentum scales $\mu \gg \Lambda_{\rm QCD}$ may be
reliably determined using perturbation theory.  Unfortunately,
it is very difficult to anticipate the size of nonperturbative
contributions in any calculation.  Warnings concerning the
enhancement of nonperturbative effects as $a\!\rightarrow\! 0$
in the presence of power divergences have been issued \cite{powdiv}.
The Lepage-Mackenzie $q^\ast$ scale (see below) can also serve
as a cautionary measure of PT's reliability.
Nevertheless, no harm is done by trying out perturbation theory
to see how well it works using as many cross-checks as possible;
in fact, much might be learned.

Good agreement between perturbative and Monte Carlo estimates
of the critical quark mass for Wilson quarks and various Creutz
ratios of Wilson loops was demonstrated in Ref.~\cite{aV}.  Estimates
of light-quark current renormalizations from boosted PT \cite{Zs1}
and nonperturbative chiral Ward identities \cite{Zs2,Zs3}
have been recently compared for the improved clover fermionic
action (without tadpole improvement).  The results are summarized
below ($Z_V$, $Z_A$, $Z_{PS}$, and $Z_S$ are the vector, axial,
pseudoscalar, and scalar current renormalizations, respectively).
\begin{center}
\begin{tabular}{rclc}
 & $\beta$ & \multicolumn{1}{c}{Chiral WI} &
 \multicolumn{1}{l}{ 1-loop BPT} \\
$Z_V$ & 6.0 & \hspace{2mm} 0.824(2) & 0.83 \\
      & 6.2 & \hspace{2mm} 0.817(9) & 0.83 \\
$Z_A$ & 6.0 & \hspace{2mm} 1.09(3)  & 0.97  \\
      & 6.2 & \hspace{2mm} 1.045(10) & 0.97 \\
$Z_{PS}/Z_S$ & 6.0 & \hspace{2mm} 0.60(2) & 0.71 \\
      & 6.2 & \hspace{2mm} 0.649(9) & 0.72
\end{tabular}
\end{center}
Other nonperturbative determinations of these currents have also
been compared to the above PT results in Ref.~\cite{vladikas}.
Perturbative estimates of the power-divergent energy shift $aE_0$
and heavy-quark mass renormalization $Z_m$ in NRQCD have
been used \cite{upsilon} to determine the $\Upsilon$ mass $aM_\Upsilon^{(a)}$
from simulation results $aE_\Upsilon$ using
$aM_\Upsilon^{(a)}=2(Z_m aM - aE_0)+aE_\Upsilon$.  The results
compare very well with the nonperturbative determinations
$aM_\Upsilon^{(b)}$ obtained by fitting to the dispersion
$aE_\Upsilon({\bf p})=aE_\Upsilon + a^2{\bf p}^2/ (2aM_\Upsilon^{(b)})$,
as shown in the table below.
\vspace{-6mm}
\begin{center}
\begin{tabular}{cccccc}
$aM$ & $aE_\Upsilon$ & $Z_m$ & $aE_0$ & $aM^{(a)}_\Upsilon$
 & $aM^{(b)}_\Upsilon$ \\
1.71 & 0.453(1) & 1.20 & 0.32 & 3.92 & 3.94(3) \\
1.80 & 0.451(1) & 1.18 & 0.31 & 4.08 & 4.09(3) \\
2.00 & 0.444(1) & 1.16 & 0.30 & 4.48 & 4.48(4)
\end{tabular}
\end{center}

\subsection{Choice of expansion parameter}
A major source of ambiguity in perturbative calculations is the
choice of expansion parameter.  Consider the perturbative prediction
for an observable $O(Q)$ depending on a single external momentum $Q$
of the generic form
\begin{eqnarray}
O(Q)\! &=&\! c_0\ \bigl[ 1
 + c_1(Q/\mu,\!{\rm RS})\  \alpha_s(\mu,\!{\rm RS}) \nonumber \\
 & &\;\;\; + c_2(Q/\mu,\!{\rm RS})\ \alpha_s^2(\mu,\!{\rm RS})
 + \cdots\bigr].
\end{eqnarray}
The coefficients $c_i(Q/\mu,\!{\rm RS})$ and the expansion parameter
$\alpha_s(\mu,\!{\rm RS})$ depend on the choice of scale $\mu$ and
the renormalization scheme (RS), {\it i.e.}, the definition of
$\alpha_s(\mu,\!{\rm RS})$, in such a way that the observable $O(Q)$ is
independent of both $\mu$ and the RS.  However, in practice, one
is forced, either by exhaustion or the asymptotic nature of the
series, to truncate the perturbation expansion at some finite
order in the coupling.  Unfortunately, the resulting finite-order
approximants are no longer independent of $\mu$ and the RS.

For a {\em good} choice of expansion parameter, the uncalculated
higher-order terms should be small.  Since these contributions are unknown,
one must ultimately rely on a guess about their sizes.  This guess can
be based on the sensitivity to changes
in the scale and the RS, the apparent rate of convergence of
the perturbation series for a variety of quantities,  the size of the
coefficient of the last term in the truncated series, or some physical
criteria.  The signal for a poor expansion parameter is the appearance
of large higher-order coefficients in the perturbative expansions of
numerous observables.

Choosing an expansion parameter involves fixing the RS (defining
the coupling), specifying the scale at which to evaluate the coupling
(in the case of a running coupling), and determining the numerical
value of the coupling at that scale.  Various examples of expansion
parameter choices are described below.

\subsubsection{Coupling definitions}
Some couplings which have appeared in the literature include:

 (a) The bare lattice coupling $\alpha_0$.  This is a poor
expansion parameter, yielding first-order corrections for various
short-distance quantities which are consistently much too small
in comparison to simulation measurements.  Large expansion
coefficients routinely appear in perturbative series when
expressed in terms of this coupling.

 (b) The ``boosted'' coupling $\alpha_b\!=\! 3\alpha_0/
\langle {\rm Tr} U_{\rm plaq}\rangle$ defined by rescaling
the bare coupling using the mean plaquette.  Here, the mean field
contributions responsible for the problems in (a)
are absorbed into the coupling, resulting in better behaved
perturbative series.

 (c) The coupling $\alpha_{SF}(q)$ defined through the Schr\"odinger
functional using a recursive finite-size scaling technique
\cite{aSFA,aSFB}.  This coupling runs with the finite system
size $L=1/q$.  An alternative coupling $\alpha_{TP}(q)$ based on
the correlations of Polyakov loops in systems with twisted boundary
conditions has also been proposed \cite{aTP}.

 (d) The coupling $\alpha_{q\bar q}(q)=\frac{4}{3}r^2 F(r)$ for
large $q=1/r$ defined in terms of the interquark force
$F(r)=dV/dr$ \cite{aqq}.

 (e) The coupling $\alpha_V(q)$ defined in terms of the short-distance
static quark potential $V(q)$ using
$V(q) \!\equiv\! - 4\pi C_F \alpha_V(q)/q^2$, for $q$ large \cite{aV}.
By absorbing higher-order
contributions to the static-quark potential into the coupling, it is
hoped that higher-order contributions to other physical quantities
in terms of this coupling will then be small.  This definition
facilitates scale setting since the running coupling's argument
can be easily related to a gluon momentum.
Order-$\alpha_V$ agreement of perturbative with simulation results
for several short-distance quantities has also been demonstrated.

 (f) The coupling $\alpha_{\overline{\rm MS}}$ defined in the
familiar modified minimal subtraction scheme.

\subsubsection{Scale settings}
Various scale setting prescriptions have been devised:

 (a) Since $\alpha_0$ and $\alpha_b$ do not run, no scale setting
is required for these couplings.

 (b) A very simple possibility is to somehow guess the scale.
For some quantities, it should be possible to crudely estimate the
scale likely to dominate the processes involved.

 (c) When the two-loop contribution is known, the scale may be
chosen so that the one-loop coefficient vanishes\cite{aSFB}.
This procedure works well when relating couplings in different
schemes (and is equivalent to choosing the relative scale as the
ratio of $\Lambda$-parameters) but has not been tested on other
quantities.

 (d) Another scale setting scheme is the
Lepage-Mackenzie $q^\ast$ prescription\cite{aV}.
For a one-loop contribution
$\int\! d^4q\ \xi(q)$, where $q$ is the momentum of the exchanged gluon,
one chooses $q^\ast$ such that $\alpha_V(q^\ast)\int d^4q\ \xi(q)
= \int d^4q\ \alpha_V(q)\ \xi(q)$.  Inserting the lowest-order form
for the running coupling into the integral on the right-hand side
yields the result $\ln(q^{\ast 2})\!=\!\int\! d^4q \ln(q^2)\ \xi(q)
/ \int\! d^4q\ \xi(q)$.  This procedure is the lattice analogue of
the Brodsky-Lepage-Mackenzie prescription in continuum perturbation
theory\cite{brod}.  Difficulties with this procedure can arise when
$\int\! d^4q\ \xi(q) \approx 0$, and note that the mean value theorem
guarantees that $q^\ast$ will satisfy $0\leq aq^\ast \leq 2\pi$
only if $\xi(q)\geq 0$ for all $q$ throughout the region of
integration (or $\leq 0$ for all such $q$).

\subsubsection{Value determinations}
Numerical values for the chosen coupling may be assigned
in various ways:

 (a) In a given simulation, determining the values for $\alpha_0$
and $\alpha_b$ is straightforward.

 (b) The value of the chosen coupling may be obtained using a
perturbative expansion in terms of the bare couplings $\alpha_0$ or
$\alpha_b$.  For example, in SU(3):
\[
\begin{array}{l}
\alpha_V(s/a)\! =\! \alpha_0 \!+\! (6.71\!-\!1.75\ln s)
 \ \alpha_0^2\! +\! \cdots,\\
\hspace{13mm} =\! \alpha_b \!+\! (2.52\!-\!1.75\ln s)\ \alpha_b^2
 \!+\! \cdots,\\
\alpha_{\overline{\rm MS}}(s/a) \!=\! \alpha_0\!+\!(5.88\!-\!1.75\ln s)
 \ \alpha_0^2 \\
\quad\quad + (43.41\!-\!21.89\ln s\!+\!3.06\ln^2\! s)\
\alpha_0^3\!+\!\cdots,\\
\hspace{15mm}=\! \alpha_b\!+\!(1.69\!-\!1.75\ln s)
 \ \alpha_b^2 \\
\quad\quad + (6.31\!-\!7.23\ln s\!+\!3.06\ln^2\! s)\ \alpha_b^3
\!+\!\cdots,\\
\alpha_{SF}(s/a) \! = \! \alpha_0 \!+\! (4.62\!-\!1.75\ln s)
 \ \alpha_0^2\! +\! \cdots,\\
\hspace{13mm} =\! \alpha_b \!+\! (0.43\!-\!1.75\ln s)\ \alpha_b^2
 \!+\! \cdots.
\end{array}
\]

 (c) The coupling strength can also be determined by accurately measuring
in a simulation some short-distance quantity whose perturbative
expansion is reliable and known to sufficiently high order.  In
Ref.~\cite{aV}, the logarithm of the mean plaquette was used:
\begin{equation}
-\!\ln\langle{\textstyle\frac{1}{3}}{\rm Tr}U_{\rm plaq}\rangle
\!=\! \textstyle{\frac{4\pi}{3}}\alpha_V(3.41/a)\left\{
1\!-\!1.185\alpha_V\right\},
\end{equation}
where the scale $3.41/a$ is determined using the $q^\ast$ prescription
mentioned above.  Values of the coupling at other large values of $q$
are then obtained using the familiar two-loop perturbative
evolution equation.

 (d) A finite-size scaling procedure can be used to measure
$\alpha_{SF}(q)$ in terms of Sommer's scale $r_0$  \cite{som}
without recourse to perturbation theory.

\section{TADPOLE IMPROVEMENT}
Tadpole improvement (TI) refers to the simple procedure of modifying
any lattice gauge field operator by rescaling
the link variable $U_\mu(x)$ by a mean field factor:
$U_\mu(x)\rightarrow U_\mu(x)/u_0$, where a convenient
gauge-invariant choice for the mean field parameter is
$u_0=\langle \frac{1}{3}{\rm Tr}U_{\rm plaq}\rangle^{1/4}$ \cite{aV}.
The purpose of this procedure is to assist in the construction of
improved lattice operators, that is, lattice operators with diminished
discretization errors and lattice-to-continuum renormalization
factors nearer to unity.

When constructing any lattice operator $O[U]$ which is a sum of various
basic gauge-field operators $O_j[U]$,
that is, $O[U]=\sum_j c_j(g)\ O_j[U]$,
where each $O_j$ is a simple gauge-invariant product of link variables,
the coefficients $c_j$ must somehow be determined.  Lattice perturbation
theory is often called upon to fix these coefficients.  However, the
$c_j$'s generally contain large mean-field contributions;
to reliably account for these tadpole effects using low-order
perturbative expansions, even when using a good expansion parameter,
is asking much from perturbation theory.  Tadpole improvement
offers a better alternative: use mean-field theory instead
of PT to treat the tadpole contributions.
By tadpole improving the basic operators,
$O[U]=\sum_j \tilde c_j(g)\ O_j[U/u_0]$,
the mean-field
effects are removed nonperturbatively, resulting in smaller leftover
coefficients $\tilde c_j$ which should be much more reliably estimated by
low-order perturbative expansions.  Note that $u_0$ is a nonperturbatively
measured parameter when simulating, but its perturbative expansion must be
used when computing the $\tilde c_j$ in PT.

Hence, tadpole improvement should be viewed as a simple means of
{\em combining} mean-field theory with perturbation
theory in order to determine the parameters in a lattice operator.
For example, consider the Symanzik-improved gluon action of
L\"uscher and Weisz \cite{glueimp} with the chair coupling set to
zero.  The ratios of the rectangle $\beta_{\rm rt}$ and
parallelogram $\beta_{\rm pg}$ couplings to the plaquette
$\beta_{\rm pl}$ coupling change under TI as follows:
\[
\begin{array}{l}
-20{\displaystyle\frac{\beta_{\rm rt}}{\beta_{\rm pl}}} = (1
+ 2.02\alpha_s) \stackrel{TI}{\longrightarrow} u_0^{-2}(1
 + 0.48\alpha_s)\\
\hspace{3mm} -{\displaystyle\frac{\beta_{\rm pg}}{\beta_{\rm pl}}}
 = 0.03\alpha_s
\stackrel{TI}{\longrightarrow} 0.03 u_0^{-2}\alpha_s,
\end{array}
\]
where $u_0$ is measured in the simulation (the basic operators
without TI correction are multiplied by these coefficients).
This is not the same as simply using the boosted $\alpha_b$.
Note that mean-field corrections are sometimes large;
heavy quarkonium spin splittings are dramatically underestimated
by a factor of $1/2$ using tree-level couplings if TI is not implemented.

\section{HIGHLIGHTS FROM RECENT PAST}
Six selected lattice perturbation theory papers or series of papers
from the recent past are surveyed in this section.  For recent
developments concerning renormalons, see Ref.~\cite{renormalon}.

\subsection{On the viability of lattice perturbation theory}
Apparent discrepancies between perturbative and Monte Carlo
estimates of various short-distance quantities were shown in
Ref.~\cite{aV} to result from the use of the bare lattice coupling
$\alpha_0$ as the expansion parameter.  An expansion parameter
$\alpha_V(q^\ast)$ defined in terms of the physical static-quark
potential was advocated.  Studying the
expectation value of the trace of a link in Landau gauge, the
critical mass for Wilson quarks, and various Creutz ratios of Wilson
loops, the authors demonstrated that such discrepancies do
not occur when a renormalized coupling such as $\alpha_V(q^\ast)$
is used.

Also in this paper, the nonlinear relation between the link operator
and the gauge field was identified as a source of large mean-field
renormalizations which hamper attempts to construct improved lattice
operators.  A tadpole improvement scheme, as previously discussed, was
suggested to remedy this problem.

Lastly, the onset of asymptotic or perturbative scaling in lattice
QCD using the standard Wilson action was investigated.
The 1P-1S mass splittings in charmonium and bottomonium
and the string tension were shown to scale well for $\beta$ values as
low as 5.7 when expressed in terms of the scale parameter $\Lambda_V$
associated with $\alpha_V$; scaling is not observed for these quantities
when expressed in terms of $\Lambda_0$ associated with the bare
lattice coupling.  This suggests that lattice spacings used in
current simulations are small enough for reliable studies of QCD.

\subsection{$\alpha_{\overline{\rm MS}}$ in terms of $\alpha_0$}
In an impressive series of papers \cite{lusA}-\cite{lusD}, L\"uscher
and Weisz have recently extended to two-loop order the perturbative
expansion of $\alpha_{\overline{\rm MS}}$ in terms of the bare lattice
coupling $\alpha_0$ for SU($N$) gauge theories.  By matching the
perturbative expansions of corresponding correlation functions
in the lattice and continuum theories, they found
\[
\alpha_{\overline{\rm MS}}(s/a) = \alpha_0 + d_1(s)\ \alpha_0^2
 + d_2(s)\ \alpha_0^3 + \cdots,
\]
where
\begin{eqnarray*}
d_1(s) &=& -\frac{11N}{6\pi}\ln s \!-\! \frac{\pi}{2N} \!+\! k_1 N,\\
d_2(s) &=& d_1^2(s) \!-\! \frac{17N^2}{12\pi^2}\ln s \!+\! \frac{3\pi^2}{8N^2}
 \!+\!  k_2 \!+\! k_3 N^2,\\
k_1 &=& 2.135730074078457(2),\\
k_2 &=& -2.8626215972(6),\\
k_3 &=& 1.24911585(3).
\end{eqnarray*}
This calculation is part of an overall strategy for measuring the
running coupling using lattice
simulations and a nonperturbatively defined coupling $\alpha_{SF}$
related to the Schr\"odinger functional.

The calculation exploits the background field technique of de Wit.
In order to place their calculation on firm theoretical grounds,
the authors first show in Ref.~\cite{lusC} that lattice gauge theory
with a background gauge field is renormalizable to all orders in
perturbation theory.  The proof is based on the BRS, background gauge,
and background shift symmetries of the lattice functional
 integral.  They find that
no new counterterms are required in addition to those already needed
in the absence of the background field.

An important advantage in using the background field technique is the
fact that the relation between $\alpha_{\overline{\rm MS}}$ and $\alpha_0$
can be extracted solely from the background field 2-point function: the
3-point vertex function need not be considered.  This dramatically
reduces the number of Feynman diagrams which must be computed.  Also,
diagrams with two external legs are much simpler to evaluate
than those with three external legs.

 The calculation is described in
detail in Ref.~\cite{lusD} and involves 4
one-loop diagrams and 31 two-loop diagrams (7 factorizable, 7 ring,
7 tadpole, 3 diamond, 3 eye, and 4 bigmac diagrams).
 All vertex factors were
generated using algebraic manipulation programs written in {\sc Maple}.
The Feynman loop-integrals were evaluated using a new, innovative
position-space method described in Ref.~\cite{lusB}.  This method
is based on efficiently evaluating the free massless propagator
in coordinate space using a recursion relation which expresses the
propagator in terms of its values close to the origin.  Convergence
of lattice sums is accelerated using known asymptotic forms of the
propagator.  The method is even useful for evaluating Feynman diagrams
with non-zero external momenta.

\subsection{Tadpole-improved heavy-light lattice operators}
In Ref.~\cite{hill}, the lattice-to-continuum renormalizations
of the temporal components of the
point $(A_\mu)$ and point-split $(A_\mu^{\rm ps})$
axial currents in the heavy-quark effective theory (HQET) were computed
to one-loop order.  The standard Wilson fermionic action with massless
light quarks was used with several values for the Wilson $r$
parameter.  This was the first HQET calculation to implement tadpole
improvement and to use $\alpha_V(q^\ast)$ as the
expansion parameter.  Writing the continuum current $A_c$ in terms of the
point and point-split lattice currents as $A_c(\mu) \approx
\tilde Z(\mu a) A_L(a)/2 \approx \tilde Z_{\rm ps}(\mu a) A^{\rm ps}_L(a)
/(2u_0)$, the renormalizations for $r=1$ were found to be
\[\begin{array}{c}
\,\,\,\,\tilde Z(\mu a) = 1 \!+\! (-1.48+0.318\ln a\mu)\,\alpha_V(2.18/a),\\
\tilde Z_{\rm ps}(\mu a) =  1\!+\! (-0.76+0.318\ln a\mu)\,\alpha_V(2.13/a).
\end{array}\]
These results may be compared to those from boosted PT in which one
writes  $A_c(\mu) \approx \sqrt{2\kappa_{bc}}
 Z(\mu a) A_L(a) \approx  \sqrt{2\kappa_{bc}} Z_{\rm ps}(\mu a)
 A^{\rm ps}_L(a)$,
where  $\kappa_{bc}=u_0\kappa_c$ is the boosted critical hopping parameter,
$\alpha_b=\alpha_0/u_0^4$, and, for $r=1$,
\begin{eqnarray*}
 Z(\mu a) &=& 1 + (-1.64+0.318\ln a\mu)\, \alpha_b,\\
 Z_{\rm ps}(\mu a) &=&  1+ (0.12+0.318\ln a\mu)\, \alpha_b.
\end{eqnarray*}
Comparisons for a few values of $\beta$ are shown in the table below:
\begin{center}
\begin{tabular}{@{}c@{\hspace{2mm}}cc@{\hspace{2mm}}cc@{\hspace{0.5mm}}c@{}}
$\beta$ & $\mu a$ & $\tilde Z/2$ & $\sqrt{2\kappa_{bc}}Z$ &
 $\tilde Z_{\rm ps}/2u_0$ & $\sqrt{2\kappa_{bc}} Z_{\rm ps}$ \\
5.7 & 1 & 0.34 & 0.40 & 0.48 & 0.55\\
6.0 & 1 & 0.37 & 0.41 & 0.49 & 0.53 \\
6.2 & 1 & 0.38 & 0.41 & 0.50 & 0.53\\
5.7 & $aq^\ast$ & 0.36 & 0.43 & 0.51 & 0.57\\
6.0 & $aq^\ast$ & 0.39 & 0.43 & 0.52 & 0.55\\
6.2 & $aq^\ast$ & 0.40 & 0.43 & 0.52 & 0.54
\end{tabular}
\end{center}
One sees that differences between the one-loop
results from tadpole-improved
renormalized PT and boosted PT can be as large as $10\%$.

\subsection{Couplings in NRQCD}
Nonrelativistic lattice QCD (NRQCD) is an effective field theory
designed for studying hadrons containing heavy quarks.  The NRQCD
action includes interactions which systematically correct for
relativity and finite-lattice-spacing errors.  The couplings
strengths of these interactions can be determined in perturbation
theory.

\begin{figure}
\begin{center}
\leavevmode
\epsfxsize=3.25 true in \epsfbox[80 460 530 750]{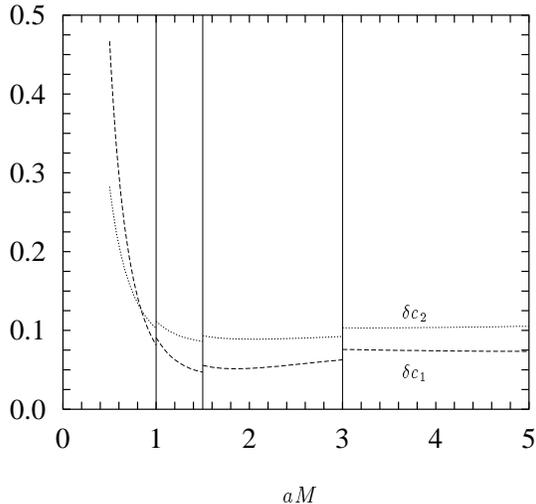}
\end{center}
\caption[nrqcd]
{The one-loop corrections $\delta c_1$ and $\delta c_2$ to two NRQCD
 couplings for $\beta=6.0$ against the bare lattice quark
 mass $aM$.
\label{NRQCDcouplings}}
\end{figure}

In Ref.~\cite{morn}, the heavy-quark mass and wave function
renormalization, energy shift, and two of the couplings in NRQCD
were calculated to leading order in perturbation theory as
functions of the bare quark mass $aM$ in lattice units.
Tadpole improvement was implemented and results were given
in terms of the QCD coupling $\alpha_V(q^\ast)$ with scales
set using the Lepage-Mackenzie $q^\ast$ prescription.  The radiative
corrections $\delta c_1$ and $\delta c_2$ to the so-called kinetic
couplings, $c_1$ and $c_2$, where $c_j=1+\delta c_j$, are shown in
Fig.~\ref{NRQCDcouplings};  $c_1$ is the coupling strength of the
$(\sum_k D_k^2)^2$ interaction, and $c_2$ is the coupling of
the $\sum_k D_k^4$ term, where $D_\mu$ is a covariant
lattice derivative.  The mass renormalization and energy
shift were used in a determination of the $b$-quark pole mass
\cite{mb}.  The complexity of the NRQCD action necessitated a
heavy reliance on symbolic manipulations using {\sc Maple} in carrying
out these calculations.

\subsection{Structure functions}
The renormalization constants and mixing coefficients for
the lowest-twist lattice operators appearing in the Wilson expansion
of the product of two hadronic currents have recently been calculated
to one-loop order by two groups in Refs.~\cite{disA,disB,disC}.
These quantities are needed to extract from simulations the hadronic
matrix elements relevant for determining moments of
the quark and gluon distributions inside hadrons,
{\it i.e.}, the deep inelastic structure functions.
The lattice operators considered were of the form
\begin{eqnarray*}
Q_{\tau_1\cdots\tau_n}^f &=& 2^{-n}
 \overline{\psi}\ \gamma_{\tau_1}
 \stackrel{\textstyle\leftrightarrow}{D}_{\tau_2}\cdots
 \stackrel{\textstyle\leftrightarrow}{D}_{\tau_n}\lambda^f\psi,\\
Q_{\tau_1\cdots\tau_n}^{5f} &=& 2^{-n}
 \overline{\psi}\ \gamma_{\tau_1}\gamma_5
 \stackrel{\textstyle\leftrightarrow}{D}_{\tau_2}\cdots
 \stackrel{\textstyle\leftrightarrow}{D}_{\tau_n}\lambda^f\psi,\\
G_{\tau_1\cdots\tau_n} &=& \sum_\rho {\rm Tr}( F_{\tau_1\rho}
 \stackrel{\textstyle\leftrightarrow}{D}_{\tau_2}\cdots
 \stackrel{\textstyle\leftrightarrow}{D}_{\tau_{n-1}}
 F_{\rho\tau_n}),
\end{eqnarray*}
ignoring trace terms, where $\psi$ is a quark field, $D_\mu$ is a
covariant lattice derivative, $F_{\mu\nu}$ is the cloverleaf gluon
field strength tensor, and $\lambda^f$ are flavour matrices.
The rank-two flavour-singlet operators $Q_{\mu\nu}$ and
$G_{\mu\nu}$ mix and are related to the first moments of the quark and
gluon distributions, respectively; the rank-three
$Q_{\mu\nu\tau}$ is related to the
second moment of the quark distribution.  The quantities calculated
were the renormalization factors
\begin{equation}
Z_{kl}(\mu a)=\delta_{kl}-\frac{\alpha_0}{4\pi}C_F\left(
\gamma_{kl}\ln \mu a + B_{kl}\right)
\end{equation}
relating the bare
lattice operators to a set of finite operators $\hat O_k(\mu)=
Z_{kl}(\mu a)O_l(a)$ renormalized by requiring that the matrix elements
of $\hat O$ in external massless quark and/or gluon states
having momentum $p^2=\mu^2$
are identical to the tree-level matrix elements of the bare lattice
operators, for $\mu a \ll 1$ to minimize discretization effects.
Only operators which cannot mix with lower dimensional
operators due to symmetry considerations were studied.

In Refs.~\cite{disA,disB}, results were obtained for both the Wilson
and the improved clover actions in the chiral limit;
tadpole improvement was not implemented.  For the rank two operators,
internal quark loops were taken into account, and
flavour singlet and non-singlet renormalizations were studied;
only the quenched theory was used for the rank three quark operators.
The calculations relied heavily on the use of the algebraic manipulation
languages {\sc Form} and {\sc Schoonschip}, with modifications to
properly treat the Dirac matrices.
The following operators were considered: $O_1=Q_{\{14\}}$,
$O_2=G_{\{14\}}$, $O_3=Q_{\{14\}}^{NS}$,
$O_A=Q_{411}-\frac{1}{2}(Q_{422}+Q_{433})$,
$O_B=Q_{141}+Q_{114}-\frac{1}{2}(Q_{242}+Q_{224}+Q_{343}+Q_{334})$,
and $O_C=Q_{\{123\}}$,
where $\{\cdots\}$ denotes symmetrization and the superscript
$NS$ denotes non-singlet (operators are flavour singlets unless
otherwise indicated).  Some selected results for $r\!=\!1$ (quenched)
are given in the table below (results for the improved fermionic
action are indicated by the superscript $I$):
\begin{center}
\begin{tabular}{ccrr}
$k,l$ & $\gamma_{kl}$  & $B_{kl}$\hspace{2mm} & $B^I_{kl}$ \hspace{2mm}\\
$1,1$ & $16/3$ & $-3.165$  &  $-15.816$ \\
$A,A$ & $13/3$ & $-18.824$ &  $-27.389$ \\
$A,B$ & $2$            & $-0.924$  &  $-3.603$ \\
$B,A$ & $4$            & $-2.955$  &  $-11.803$ \\
$B,B$ & $19/3$ & $-17.540$ &  $-18.538$ \\
$C,C$ & $25/3$ & $-19.005$ &  $-29.815$
\end{tabular}
\end{center}
The rank two renormalizations using the Wilson
action were compared with previous determinations:  some
discrepancies were found.

The Wilson action results above were confirmed in Ref.~\cite{disC}
(except the last row which was not computed).
However, these authors advocate the use of the operators
$\tilde O_A=O_A+O_B$ and $\tilde O_B=O_B-2O_A$ in order to
diagonalize the anomalous dimension matrix.
They also considered the following operators: $O_{1b}=Q_{44}-
\frac{1}{3}(Q_{11}+Q_{22}+Q_{33})$,
$O_4=Q^5_2$, $O_5=Q^5_{\{214\}}$, and $O_6=Q^5_{[2\{1]4\}}$,
where $[\cdots]$ denotes antisymmetrization.  The
renormalizations using the $r=1$ Wilson action are given below:
\begin{center}
\begin{tabular}{ccr}
$k$  & $\gamma_{kk}$  & $B_{kk}$\hspace{5mm} \\
$1b$ & $16/3$ & $-1.892(06)$ \\
$4$  & $0$    & $15.795(03)$ \\
$5$  & $25/3$ & $-19.560(10)$ \\
$6$  & $7/3$  & $-15.680(10)$
\end{tabular}
\end{center}
{\sc Mathematica} and {\sc Maple} were used to perform
the calculations.

\subsection{Stochastic perturbation theory}
Recently, an innovative numerical technique \cite{stochA}
for obtaining weak-coupling perturbative expansions of local observables
in lattice QCD was proposed.  An exciting aspect of this method is
that it allows one to obtain much longer expansions than
presently possible using conventional diagrammatic approaches.

The method is based
on Parisi-Wu stochastic quantization in which the gauge field is
viewed as a random variable which evolves according to the
Langevin equation.  One step $t\!\!\rightarrow\!\! t\!+\!\epsilon$
in the discrete Langevin equation
consists of a sweep through the lattice, updating links
using $U_\mu(x;t\!+\!\epsilon)\!=\!\exp[-F_\mu(x;t)]U_\mu(x;t)$, where the
driving function $F_\mu$ depends on the link variables in all plaquettes
containing the link between sites $x$ and $x\!+\!a\hat\mu$,
 the Langevin time step
$\epsilon$, and a noise matrix.  One then writes $U_\mu(x;t)=
\exp[A_\mu(x;t)/\sqrt{\beta}]$, rescales the time step
$\epsilon=\tau/\beta$, expands $A_\mu(x;t)=\sum_{k\geq 0}
\beta^{-k/2}\ A_\mu^{(k)}(x;t)$ and $F_\mu$ as power series in
$1/\sqrt{\beta}$, and truncates to some order to transform the
Langevin equation into a system of coupled stochastic finite-difference
equations.  The coefficients of the perturbative expansion of any
local observable $W$ are then given by expectation
values of composite operators of the $A_\mu^{(k)}$ which are
obtained by averaging over the Langevin history:
\begin{eqnarray}
W &=& {\textstyle\sum_n} \beta^{-n/2}\langle O_n\rangle,\\
\langle O_n\rangle &=&  \lim_{T\rightarrow\infty} \frac{1}{T}
 \sum_{t=1}^T O_n(t).
\end{eqnarray}
The discrete Langevin equation has $O(\epsilon)$ systematic errors,
making it necessary to extrapolate the results to $\epsilon\rightarrow 0$.
Stochastic gauge fixing is necessary for an acceptable signal-to-noise
ratio and finite size errors must be controlled.

In Ref.~\cite{stochB}, the mean plaquette in SU(3),
\begin{equation}
 P = 1-\frac{1}{3}\langle {\rm Tr}\ U_p\rangle =
\sum_{n=1}^\infty c_n\ \beta^{-n},
\end{equation}
 was computed to {\em eight}-loop order.
Results were obtained on an $8^4$ lattice with periodic boundary
conditions for step sizes $\tau=0.01, 0.015, 0.02$,
then extrapolated to $\tau\rightarrow 0$.  The first three coefficients
were found to agree with known values obtained analytically
in Ref.~\cite{alles}
using the standard diagrammatic approach, as shown in the table below.
\begin{center}
\begin{tabular}{ccc}
 & Langevin & Analytic \\
$c_1$ &  1.998(1) & 2        \\
$c_2$ &  1.218(1) & 1.212(7) \\
$c_3$ &  2.940(5) & 2.9605   \\
$c_4$ &  9.28(2) & \\
$c_5$ &  34.0(2) & \\
$c_6$ &  134.9(9) & \\
$c_7$ &  563(5) & \\
$c_8$ &  2488(29) &
\end{tabular}
\end{center}

\section{CONCLUDING REMARKS}
The evaluation of Feynman diagrams in lattice PT is difficult
because the Feynman integrands are usually complicated
functions of the loop and external momenta.  Standard tools, such as
Feynman parameters and partial integration methods, are
not very helpful on the lattice.  Progress is being made,
however, with the development of new techniques, such as those proposed
in Refs.~\cite{lusA}-\cite{lusD}, an increasing reliance on and expertise
in using analytical computer programs, such as {\sc Maple} and
{\sc Mathematica}, and the introduction of exciting new stochastic
methods.  Lattice perturbation theory continues to evolve and to play
an important role in lattice field theory.

I thank S.~Capitani, C.~Davies, G.P.~Lepage, D.~Richards,
G.~Schierholz, J.~Shigemitsu, Z.~Weihong, and U.~Wolff for useful
discussions, and acknowledge the financial support of PPARC
through grant GR/J 21347.

\end{document}